\documentstyle[twoside,fleqn,espcrc2,epsf]{article}
\epsfverbosetrue
\title{\vspace{-.5in}\hbox{}\hfill{\normalsize ANL-HEP-CP-98-65}\\
\vspace{.3cm}
Domain wall fermions at finite temperature.}
 
\author{
J.-F. Laga\"e\address {HEP Division, Argonne National Laboratory, 
                       9700 South Cass Avenue, Argonne IL 60439, USA}%
\thanks{Supported by DOE contract W-31-109-ENG-38.} 
and D.K. Sinclair$\null^{{\rm a}*}$}

\begin{document}
 
\begin{abstract}
We investigate the properties of domain wall fermions on a set of quenched 
configurations at non-zero temperature. In particular, we compute the low 
lying eigenvalues of the DWF operator and study their relation with topology, 
level crossings and chiral symmetry breaking. We also measure the screening 
correlators and discuss the dependence on the extent of the extra dimension 
and the quark mass.
\end{abstract}
 

\maketitle

\section{Introduction}

An interesting aspect of the Domain Wall Fermion (DWF) formulation is that 
it provides a consistent framework for studying the chiral limit $m_q 
\rightarrow 0$. In particular, it is expected to satisfy the Atiyah-Singer 
index theorem. This is a distinct advantage over the Wilson and staggered 
formulations which are well known to suffer from shifts of the low-lying 
eigenvalues in the real and imaginary directions respectively \cite{VINK}. 
However, 
the DWF formulation accomplishes this feat at the expense of introducing 
a $5^{th}$ dimension and, strictly speaking, all the nice properties are 
only recovered in the limit $N_5 \rightarrow \infty$ (overlap formalism). 
A careful (non-perturbative) study of the convergence is therefore needed 
before practical simulations can be carried out. In this study, we initiate 
such a program in the context of QCD at non-zero temperature (thereby 
continuing our effort to understand the link between topology and the 
chiral phase transition \cite{KLS}). We have used for our computations a set of 
quenched configurations on a $16^3 \times 8$ lattice. The $\beta$ values 
[and number of configurations] studied are: 6.2[170], 6.1[170], 6.0[100],
5.9[100] and 5.8[100]. These configurations were used earlier to compute 
the screening correlators and the low-lying eigenvalue spectrum with a 
staggered fermionic action \cite{KLS}. Here, we first study the level 
crossings of the hermitian Wilson operator (in a manner similar to \cite{EHN}), 
then move on to 
compute the spectrum of the DWF operator (also properly hermitized). The 
DWF calculations are carried out (so far) at $N_5 = 4$, 6, 8 and 10. The 
link between topology, the number of level crossings and the number of DWF 
eigenvalues which vanish (exponentially) wiht $N_5$ can then be studied. 
As was noted earlier for staggered fermions \cite{KLS}, the knowledge of the 
low-lying eigenvectors is also quite valuable since in the high temperature 
phase, their contributions dominate the disconnected mesonic correlators. 
We confirm this again for DWF and use it to improve the convergence of our 
conjugate gradient inverter and to lower the error bars on the disconnected 
correlators.

\begin{figure}[htb]
\epsfxsize=7.5cm
\epsffile[70 70 550 550]{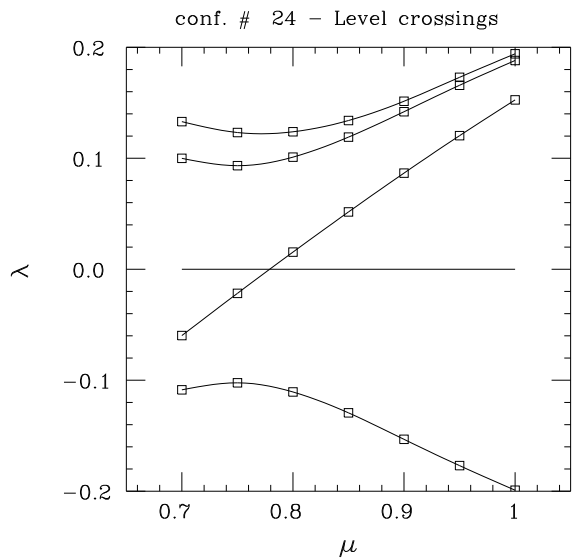}
\vspace{-1.2cm}
\caption{ Level crossing of the hermitian Wilson operator on a configuration 
with topological charge 1.}
\vspace{-0.0cm}
\end{figure}

\begin{figure}[htb]
\epsfxsize=7.5cm
\epsffile[70 70 550 550]{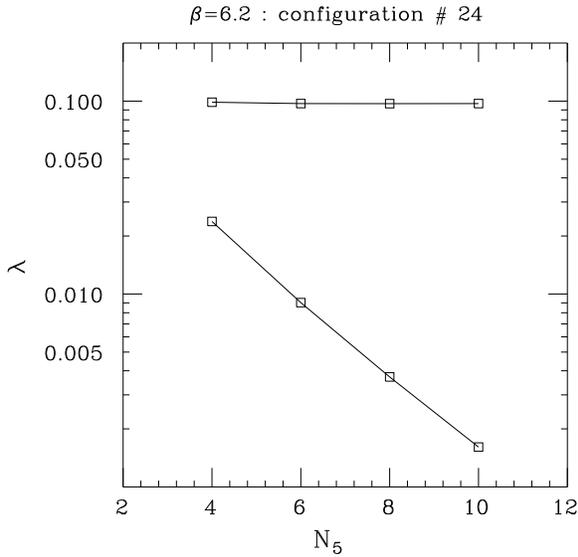}
\vspace{-1.2cm}
\caption{ Eigenvalues of the hermitian DWF operator vs. $N_5$ on a 
configuration with $Q_{top}=1$.}
\vspace{-0.2cm}
\end{figure}

\section{Results}

In this preliminary report, we focus our attention on the results obtained 
at $\beta=6.2$ (the highest temperature we have studied). The complete results, including the trends observed as one lowers the temperature towards the 
chiral phase transition will be reported elsewhere \cite{paper}. 
Also because of lack 
of space we illustrate the various computations on a single configuration. 
We have selected a configuration which has topological charge 1. We then 
obtain one level crossing in the spectrum of the hermitian Wilson operator 
(Fig. 1). It occurs at $\mu \approx 0.78$ with a slope of $0.74$. On our 
sample of configurations, most crossings occured between $\mu=0.75$ and 
$\mu=0.80$. One notable exception was the single configuration of topological 
charge 2 for which the second crossing occured at $\mu \simeq 0.95$ . There 
is a relationship between the presence of these crossings and the existence of 
zero modes of the hermitian operator $H \equiv \gamma_5 R D$ (where D is the 
DWF operator and R is the reflection operator in the $5^{th}$ dimension).
For example, a crossing at $\mu$ with slope $\pm 1$, would imply an exact 
zero-mode of H at a domain wall mass $M = 1 + \mu$. Around this value and 
for slopes of magnitude less than 1, there will be near zero-modes. In this 
study we have fixed $M=1.7$ (similarly to what has been done in zero 
temperature applications \cite{BLUM}). On the same configuration as before, 
We then find that the lowest eigenvalue of H decreases exponentially with 
$N_5$, whereas the second lowest is essentially independent of $N_5$ (Fig. 2). 
The lowest eigenvalue is of order 0.0016 at $N_5=10$ 
and would be 2 order of magnitudes smaller if $N_5$ was extended to 20.

\begin{figure}[htb]
\epsfxsize=7.5cm
\epsffile[70 70 550 550]{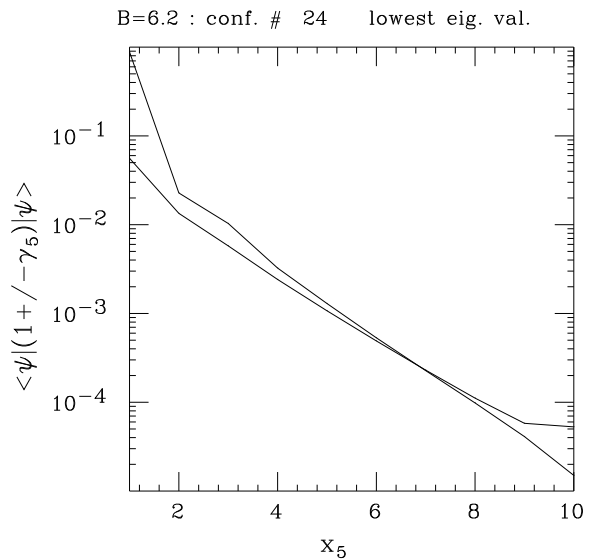}
\vspace{-1.2cm}
\caption{ 4-dimensional slices of the ``Left'' and ``Right'' components of the 
lowest eigenmode }
\vspace{-0.2cm}
\end{figure}

Our variational method for computing the eigenvalues also gives us the 
eigenvectors. Fig. 3 represents the positive and negative chirality 
components measured on the lowest mode as a function of $x_5$. The negative 
chirality component is peaked on the left wall (as should be for a 
configuration with $Q_{top}=-1$) and decreases exponentially in the interior. 
The right chirality component also reaches its maximum on the left wall but 
with a smaller magnitude. For modes with no net chirality, one would 
obtain a symmetrical graph with each component peaked on their respective 
wall. (This is what is observed for the ``first excited state'' on our sample 
configuration and also for the lowest mode on configurations with 
$Q_{top} = 0$). On a few ``abnormal'' configurations with very late crossings 
\cite{EHN}, we found eigenmodes with a net chirality but an eigenvalue which 
remains large at $N_5=10$. A detailed discussion of this (small) artifact is 
left for \cite{paper}.

Finally, as an introduction to our computation of disconnected 
mesonic correlators \cite{paper}, we present our results for the measurement of 
m $Tr \gamma_5 S$ as  a function of m at $N_5=10$. The continuum answer should 
be equal to the topological charge of the configuration ( -1 in our case ).
At low values of the quark mass,  fig.4 shows that the measurements dip towards 0. This is as expected for a situation where the lowest eigenvalue is not 
quite 0 yet at $N_5 = 10$. It is also important to note that a precise 
measurement of this operator was made possible by ``separating out'' the 
contribution from the lowest eigenmodes in the conjugate gradient inversion. 
Good results could then be obtained by using only 10 noise vectors on each 
wall. Without this ``trick'', even increasing the number of noise vectors by 
a factor of 10, still leaves us with rather large error bars (see Fig. 5).

\begin{figure}[htb]
\epsfxsize=7.5cm
\epsffile[70 70 550 550]{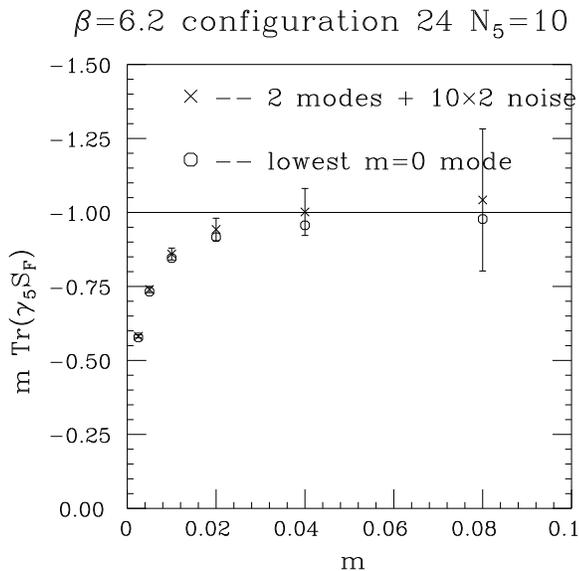}
\vspace{-1.2cm}
\caption{ Chirality versus quark mass (using our knowledge of the low-lying 
modes) }
\vspace{-0.2cm}
\end{figure}

\section{Conclusions}

We have confirmed (non-perturbatively) that DWF provide a practical and 
systematically improvable way of studying the topological properties of QCD 
(at least in the high temperature phase). The ``rate of convergence'' at 
$\beta=6.2$ can be read off from Fig.2 (and similar results at lower $\beta$ 
will be presented elsewhere). In addition, we have shown how the knowledge 
of the low-lying eigenmodes can be used to greatly improve the quality of 
fermionic measurements on configurations with non-trivial topology. We also
expect that enhancements of this technique will play an important role in 
the design of dynamical fermion algorithms at very low quark masses.

\begin{figure}[htb]
\epsfxsize=7.5cm
\epsffile[70 70 550 550]{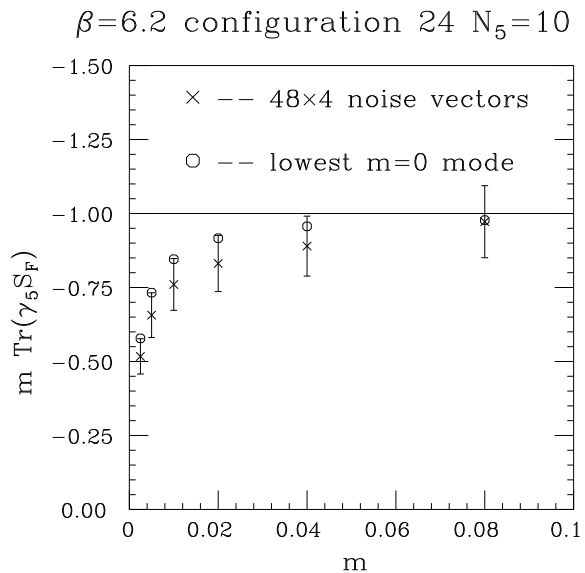}
\vspace{-1.2cm}
\caption{ Chirality versus quark mass (noisy estimator only) }
\vspace{-0.2cm}
\end{figure}

\end{document}